\documentclass[aps,11pt,superscriptaddress,preprintnumbers,
                secnumarabic,nofootinbib]{revtex4}

\usepackage{epsfig}
\usepackage{bbm}

\begin{document} 

\title{\Large Lamb Shift of Unruh Detector Levels}

\author{Bj\"orn~Garbrecht}
\email[]{Bjorn@HEP.Man.Ac.UK}
\affiliation{School of Physics \& Astronomy, University of Manchester, Oxford Road, Manchester M13 9PL, United Kingdom}

\author{Tomislav Prokopec}
\email[]{T.Prokopec@Phys.UU.NL}
\affiliation{Institute for Theoretical Physics (ITF) \& Spinoza Institute, Utrecht University, Leuvenlaan 4, Postbus 80.195, 3508 TD Utrecht, The Netherlands}

\preprint{SPIN-05/25}
\preprint{ITP-UU-05/31}

\begin{abstract}
We argue that the energy levels of an Unruh detector experience an effect
similar to the Lamb shift in Quantum Electrodynamics. As a consequence, the
spectrum of energy levels in a curved background is different from that
in flat space. As examples, we consider a detector in an expanding Universe
and in Rindler space, and for the latter case we suggest a new expression
for the local virtual energy density seen by an accelerated observer.
In the ultraviolet domain, that is when the space between
the energy levels is larger than the Hubble rate or the acceleration of
the detector, the Lamb shift quantitatively dominates over the thermal
response rate.
\end{abstract}


\maketitle

\section{Unruh Detector as a Probe for the Quantum Vacuum}

In order to gain an understanding of the physical effects which
occur in curved spacetimes, the Unruh detector~\cite{Unruh:1976}
is often considered.
It is an idealized point-like device, solely defined through its energy
levels $E_m$ by
\begin{equation}
|m,\tau\rangle={\rm e}^{-{\rm i}H_D \tau}|m\rangle
={\rm e}^{-{\rm i}E_m \tau}|m\rangle\,,
\end{equation}
where $H_D$ is the unperturbed detector Hamiltonian and $\tau$ denotes the
proper time along the detector trajectory.

Interactions with a scalar field $\phi$ take place \emph{via} the
perturbation Hamiltonian
\begin{equation}
\delta H=\hat{h}\phi(x)\,,
\label{hamilton:interaction}
\end{equation}
and we denote the matrix elements of the operator $\hat h$ as
$h_{mn}=\langle m | \hat h | n \rangle$. Let $P_{mn}(\tau)$ be the probability
for a transition $m \rightarrow n$ after the proper time $\tau$ has elapsed,
and define ${\cal F}_{mn}=P_{mn}/|h_{mn}|^2$. Then, according to the
quantum mechanical rules of time-dependent perturbation theory, the
\emph{response function} for $\tau \rightarrow \infty$ is given
by~\cite{Unruh:1976,GarbrechtProkopec:2004:1},
\begin{equation}
\label{ResponseFunction}
\frac{d{\mathcal{F}}(\Delta E)}{d\tau}
   =\int_{-\infty}^\infty d{\Delta \tau} {\rm e}^{{\rm i}\Delta E \Delta \tau}
      \langle i|
          \phi\left(x(-\Delta \tau/2)\right)\phi\left(x(\Delta \tau/2)\right)
       |i\rangle
\,,
\label{response function}
\end{equation}
where $\Delta E= E_n - E_m$, and $|i\rangle$ denotes the state of
the $\phi$-field.

This response function is often discussed for a detector in de Sitter
space~\cite{GibbonsHawking:1977}.
We describe the scalar field $\phi$ by the Lagrangean
\begin{equation}
\label{Lagrangian:scalar}
\sqrt{-g}\mathcal{L}
  =\sqrt{-g}\left(\frac{1}{2}g_{\mu\nu}\partial^\mu\phi\partial^\nu\phi
-\frac{1}{2}m^2\phi^2-\frac{1}{2}\xi R\phi^2\right)\,,
\end{equation}
implying the Euler-Lagrange equation of motion
\begin{equation}
\label{KleinGordon}
\left[\nabla^2+m^2+\xi R\right] \phi(x)=0\,,
\end{equation}
where $\nabla$ denotes the covariant derivative, such that
$\nabla^2 = (-g)^{-1/2}\partial_\mu g^{\mu\nu}\sqrt{-g}\partial_\nu$.
The spatially flat Friedmann-Lema\^itre-Robertson-Walker (FLRW) Universe
has the metric tensor
\begin{equation}
g_{\mu\nu}=a^2(\eta)\times{\rm diag}(1,-1,-1,-1)\,,
\end{equation}
where $\eta$ is the conformal time. For de Sitter space expanding at the Hubble
rate $H$, the scale factor as a function of conformal time is
\begin{equation}
\label{ScaleFactor:deSitter}
a(\eta)=-\frac{1}{H\eta}\,.
\end{equation}


We substitute $\varphi=a\phi$ and decompose $\varphi$ in modes as
\begin{equation}
\label{varphi}
  \varphi(x) = \int \frac{d^3 k}{(2\pi)^3}
              \Big(
                   {\rm e}^{i\mathbf{k}\cdot\mathbf{x}}
                    \varphi(\mathbf{k},\eta)
                            a({\mathbf{k}})
                +  {\rm e}^{-i\mathbf{k}\cdot\mathbf{x}}
                     \varphi^*(\mathbf{k},\eta)
                            a^\dagger({\mathbf{k}})
              \Big)
\,.
\end{equation}
Here $a({\mathbf{k}})$ and $a^\dagger({\mathbf{k}})$ denote the 
annihilation and creation operators for the mode with 
a comoving momentum $\mathbf{k}$, and they are defined by 
$a^\dagger({\mathbf{k}})|0\rangle = |\mathbf{k}\rangle$,
$a({\mathbf{k}})|\mathbf{k}^\prime\rangle 
  = (2\pi)^3\delta^3(\mathbf{k}-\mathbf{k}^\prime)|0\rangle$,
where $|0\rangle$ denotes the vacuum state
and $|\mathbf{k}\rangle$ the one-particle state with 
momentum $\mathbf{k}$. The physical momentum is related to the conformal
momentum by $\mathbf{k}_{\rm{phys}}=\mathbf{k}/a(\eta)$.
From the Klein-Gordon equation~(\ref{KleinGordon}), one can derive
for the mode functions $\varphi(\mathbf{k},\eta)$ the following equation:
\begin{equation}
\Big(\partial_\eta^2+\big(\mathbf{k}^2+a^2m^2\big)
   + (6\xi-1)\frac{a''}{a}
 \Big)
\varphi(\mathbf{k},\eta)=0
\label{Klein-Gordon:Quadratic}
\,,
\end{equation}
and spatial homogeneity implies
$\varphi(\mathbf{k}) =\varphi(k)$ ($k \equiv |\mathbf{k}|$). We define
the vacuum by the choice of solutions such that $\varphi(\mathbf{k},\eta)$
reduces to a plane wave of purely negative frequency at infinitely early
times.
The field $\varphi$ obeys the canonical commutation relation
\begin{equation}
[\varphi(\mathbf{x},\eta),\partial_\eta\varphi(\mathbf{x'},\eta)] 
= {\rm i}\delta^3(\mathbf{x}-\mathbf{x'}),
\label{canonical-commutation}
\end{equation}
\noindent
which implies the normalisation of the mode functions by the Wronskian
\begin{equation}
  \varphi(\mathbf{k},\eta){\varphi^*}^\prime(\mathbf{k},\eta)
 -\varphi^{\prime}(\mathbf{k},\eta)\varphi^*(\mathbf{k},\eta) = {\rm i}\,,
\label{Wronskian}
\end{equation}
and for the creation and annihilation operators the commutator
\begin{equation}
[a_\mathbf{k},a^\dagger_\mathbf{k'}] 
 = (2\pi)^3\delta^{(3)}(\mathbf{k}-\mathbf{k'})\,.
\end{equation}

The explicit solution of the mode equation~(\ref{Klein-Gordon:Quadratic})
for the minimally coupled massless case, $m=0$ and $\xi=0$,
in de Sitter space with the scale factor~(\ref{ScaleFactor:deSitter}) is
\begin{equation}
\label{phi:dS}
\varphi(\mathbf{k},\eta)=\frac{1}{\sqrt{2k}}\left(1-\frac{\rm i}{k\eta}\right)
{\rm e}^{-{\rm i} k \eta}\,,
\end{equation}
from which, when deriving the response function, one
finds~\cite{Higuchi:1986,GarbrechtProkopec:2004:1}
\begin{equation}
\label{ResponeFunction:m=0}
\frac{d{\cal F}(\Delta E)}{d \tau}=
\frac{\Delta E}{2\pi}\bigg(\!1+\!\frac{H^2}{\Delta E^2}\bigg)
                   \frac{1}{{\rm e}^{(2\pi/H)\Delta E}-1}
\quad\textnormal{for}\quad \Delta E\not=0
\,.
\end{equation} 
This result indicates an exponentially falling spectrum of scalar quanta in
de Sitter space, often interpreted as the presence of thermal radiation.

On the other hand, when calculating the energy density from the
unrenormalised stress energy tensor, we obtain~\cite{GarbrechtProkopec:2004:2}
\begin{equation}
\label{enden:dS}
\varrho=
\langle0|T^0_{\phantom{0}0}(x)|0\rangle = \frac{1}{a^4}
 \int \frac{d^3k}{(2\pi)^3}
\left(k+\frac{1}{2k\eta^2}\right)\,,
\end{equation}
suggesting that besides the cosmological constant contribution
corresponding to the first term of the integrand, there
additionally is a particle
spectrum $\propto 1/k^2$ produced by the de Sitter background, which is
apparently not captured by the response rate~(\ref{ResponeFunction:m=0}).
A similar power-law behaviour of the stress energy tensor holds
more generally also
for massive or nonminimally coupled scalars as well as for adiabatically
expanding spacetimes, such as matter and radiation
Universes~\cite{GarbrechtProkopec:2004:2,ParkerFulling:1974,HabibMolina-ParisMottola:1999}.
Since we are dealing with effects of the quantum vacuum in curved space,
in order to understand this discrepancy,
it may be useful to recall what possibly related phenomena there are in
flat space.

The first experimental result to find an explanation by vacuum
fluctuations was the Lamb shift. According to relativistic quantum
mechanics, the energy levels $2S_{1/2}$ and $2P_{1/2}$ of hydrogen 
are degenerate, despite a tiny correction due to the hyperfine structure,
insufficient however to account for the actual shift, which was observed
by Lamb and Retherford~\cite{LambRetherford:1947} in 1947.
Also in 1947, Bethe has shown in a groundbreaking paper~\cite{Bethe:1947} that
the split is due to
interactions of the electron with the vacuum fluctuations of the
electromagnetic field, and a finite answer is obtained when subtracting
the self-energy corrections for a free electron, which are infinite,
from those of an electron
in the Coulomb potential. This is probably the most illustrative, simple and
beautiful example for the effects of the quantum vacuum,
detected by the hydrogen atom as a probe. 

Just like an atom, Unruh's detector is a system with discrete energy
levels, which by Bethe's argument should also acquire a Lamb shift correction
from the fluctuations of the scalar field $\phi$. Since quantum field
theory in curved space deals with the distortions of the quantum
vacuum induced by the gravitational background, it is perhaps more
natural to expect that these become manifest in the Lamb shift rather
than in the detection rate of scalar quanta.

Therefore, we calculate in the following the self-energy corrections to the
energy levels of an Unruh detector in a spacetime $X$.
At second order in perturbation theory, these are given by~\cite{Bethe:1947}
\begin{eqnarray}
\label{LambShift}
\delta E_{n\,X}\!\!\!&=&\!\!\!\sum\limits_{m\not=n}\int\frac{d^3k}{(2\pi)^3}
\frac{\left|
\int \frac{d^3k^\prime}{(2\pi)^3} \langle \mathbf{k}^\prime,m|
\hat h a^\dagger(\mathbf{k})\varphi(\mathbf{k},\eta)
|0,n\rangle\right|^2}
{E_n-E_m-\Omega(\mathbf{k})}\\
\!\!\!&=&\!\!\!\sum\limits_{m\not=n}\int\frac{d^3k}{(2\pi)^3}
\frac{\left|h_{mn}^2 \right|\left|\varphi(\mathbf{k},\eta)\right|^2}
{E_n-E_m-\Omega(\mathbf{k})}
\,,
\nonumber
\end{eqnarray}
where $\Omega(\mathbf{k})$ is the canonical Hamiltonian
energy~(\ref{ModeEnergy}) of a $\phi$ quantum at momentum $\mathbf{k}$.
This shift of energy levels has in flat space a square divergence in the
ultraviolet. In Minkowski space, one can determine the values of 
the detector's energy levels $E_n$, which are finite, by observation
and the infinite shift $\delta E_{n\,\rm M}$ is already taken into account
for this measurement.
In a curved spacetime $C$ however, the value for the radiative correction
differs from the Minkowski space answer; the finite quantity
\begin{equation}
\label{LambShift:Renormalized}
\delta E_n=\delta E_{n\,C}-\delta E_{n\,\rm M}
\end{equation}
can therefore be observed by comparing the spectra of energy levels
in flat and in curved background.

To keep notation simple, we drop the summation over energy levels in the
following,
corresponding to a two-level detector with spacing $\Delta E\equiv E_n-E_m$
and $|h_{mn}|^2\equiv h^2$. The sum can
simply be reinserted into all subsequent results.

\section{Lamb Shift in the Expanding Universe}
The Hamiltonian for the scalar field in the FLRW-background is
given by~\cite{GarbrechtProkopecSchmidt:2002}
\begin{eqnarray}
\label{Hamiltonian:bosons}
H(\eta) =\! \frac12\!\int\!\! \frac{d^3k}{(2\pi)^3}\!\left\{
      \Omega(\mathbf{k},\eta)(a(\mathbf{k})a^{\dagger}(\mathbf{k})
   \!+\!  a^{\dagger}(\mathbf{k})a(\mathbf{k}))
   \!+\!  (\Lambda(\mathbf{k},\eta) a(\mathbf{k})a(-\mathbf{k})
   \!+\! {\rm h.c.})\!\right\},
\end{eqnarray}
where
\begin{eqnarray}
\Omega(\mathbf{k},\eta)
    \!\!\!&=&\!\!\! \left|\varphi^\prime(\mathbf{k},\eta) - (a'/a)\varphi(\mathbf{k},\eta)
        \right|^2
     + \bar\omega^2(\mathbf{k},\eta) \left|\varphi(\mathbf{k},\eta)\right|^2,
\label{ModeEnergy}
\\
\Lambda(\mathbf{k},\eta)
  \!\!\! &=&\!\!\! \Bigl(\varphi^\prime(\mathbf{k},\eta)
            -\frac{a'}{a}\varphi(\mathbf{k},\eta)\Bigr)^2
    +  \bar\omega^2(\mathbf{k},\eta)\varphi^2(\mathbf{k},\eta)
\,,
\end{eqnarray}
and we have defined $\omega^2(\mathbf{k},\eta)=\mathbf{k}^2+a^2m^2$
and $\bar\omega^2(\mathbf{k},\eta) =
\omega^2(\mathbf{k},\eta)+ 6\xi \frac{a^{\prime\prime}}{a}$.

The quantity $\Omega(\mathbf{k})$ therefore is the vacuum expectation value
for the Hamiltonian energy of the $\phi$-mode at momentum $\mathbf{k}$.
Note that this canonical energy density equals the covariant
energy density as obtained from the stress-energy
tensor~\cite{GarbrechtProkopec:2004:1,ParkerFulling:1974,HabibMolina-ParisMottola:1999} only in the minimally coupled case $\xi=0$.

We now make use of these expressions to calculate the shift of detector levels.

\subsection{Massless de Sitter Case}

As first example, let us consider a minimally coupled massless scalar
in de Sitter space because for this situation, exact solutions are
available and we do not need to resort to approximation by
adiabatic expansion.
First, we calculate the shift in Minkowski space,
\begin{eqnarray}
\delta E_{\rm M}^{m=0}\!\!\!&=&\!\!\!\int \frac{d^3k}{(2\pi)^3}\frac{1}{2k}
\frac{h^2}{\Delta E -k}
=\frac{h^2}{4\pi^2}\int\limits_0^{\infty}dk \frac{k}{\Delta E -k}\\
\!\!\!&=&\!\!\!\frac{h^2}{4\pi^2}\int\limits_0^{\infty}dk
\left\{
-1 + \frac{\Delta E}{\Delta E -k}\right\}
=\frac{h^2}{4\pi^2}
\left[-k-\Delta E \log(\Delta E -k)\right]_0^\infty
\nonumber
\,,
\end{eqnarray}
which is divergent and to be subtracted.

The de Sitter mode functions are given by Eqn.~(\ref{phi:dS}), such that
we find for their squared amplitude
\begin{equation}
|\varphi(\mathbf{k},\eta)|^2=\frac{1}{2k}+\frac{1}{2k^3\eta^2}
\end{equation}
and for the mode energy~(\ref{ModeEnergy})
\begin{equation}
\Omega(\mathbf{k},\eta)=k+\frac{1}{2k\eta^2}\,.
\end{equation}
Let us fix $a=1$, such that conformal and physical momentum coincide
and we have by Eqn.~(\ref{ScaleFactor:deSitter}) $\eta=-H^{-1}$.
The use of Eqn.~(\ref{LambShift}) gives us the unrenormalized Lamb shift
in de Sitter space,
\begin{eqnarray}
\delta E_{\rm dS}^{m=0}\!\!\!&=&\!\!\!\int \frac{d^3k}{(2\pi)^3}
\left(\frac{1}{2k}+\frac{H^2}{2k^3}\right)
\frac{h^2}{\Delta E -\left(k+\frac{H^2}{k}\right)}\\
\!\!\!&=&\!\!\!\frac{h^2}{4\pi^2}\int\limits_0^{\infty}dk
\left\{
-1 + \frac{\Delta E}{\Delta E -\left(k+\frac{H^2}{k}\right)}\right\}
\nonumber\\
\!\!\!&=&\!\!\!
\frac{h^2}{4\pi^2}\left\{
\left[-k\right]_0^\infty
-\Delta E \int\limits_{-\Delta E/2}^\infty dl
\frac{l+\Delta E /2}{l^2+H^2-\Delta E^2/4}\right\}
\nonumber\\
\!\!\!&=&\!\!\!
\frac{h^2}{4\pi^2}
\left[
-k + \frac{\Delta E^2/4}{\sqrt{\Delta E^2/4-H^2}}
\log\left|
\frac{k-\Delta E/2+\sqrt{\Delta E^2/4-H^2}}
{k-\Delta E/2-\sqrt{\Delta E^2/4-H^2}}
\right|
\right.
\nonumber\\
&&
\left.
-\frac{\Delta E}{2}\log\left|
\frac{\left(k+\Delta E/2\right)^2}{\Delta E^2/4 - H^2}-1
\right|
\right]_0^\infty
\nonumber\,.
\end{eqnarray}
We evaluate the boundary terms and subtract the flat space result to find
for the finite observable shift~(\ref{LambShift:Renormalized})
\begin{eqnarray}
\label{LambShift:deSitter}
\delta E\!\!\!&=&\!\!\!\delta E_{\rm dS}^{m=0}-\delta E_{\rm M}^{m=0}\\
\!\!\!&=&\!\!\!\frac{h^2}{4\pi^2}\left\{
\Delta E \log\left|\frac{H}{\Delta E}\right|
-\frac{\Delta E^2}{4\sqrt{\Delta E^2/4-H^2}}
\log\left|\frac{\Delta E/2 - \sqrt{\Delta E^2/4-H^2}}
               {\Delta E/2 + \sqrt{\Delta E^2/4-H^2}}\right|
\right\}.
\nonumber
\end{eqnarray}
This expression condenses considerably when expanded in $H/\Delta E$:
\begin{equation}
\delta E=\frac{h^2}{4\pi^2}\frac{H^2}{\Delta E}\left(-1 
-2\log\left|\frac{H}{\Delta E}\right|
+O\left(\frac{H}{\Delta E}\right)
\right),
\end{equation}
and when we reintroduce the sum to treat the case of more than two energy
levels, it reads
\begin{equation}
\delta E=\sum\limits_{m\not=n}
\frac{|h_{mn}|^2}{4\pi^2}\frac{H^2}{E_n-E_m}\left(-1
-2\log\left|\frac{H}{E_n-E_m}\right|
+O\left(\frac{H}{E_n-E_m}\right)
\right)\,.
\end{equation}

When compared to
the response function in de Sitter~(\ref{ResponeFunction:m=0}), which decays
exponentially in $\Delta E$, this power law behaviour becomes more important
in the ultraviolet. Since the mode energy $\Omega$ is contributing,
we can consider Lamb shift as a way to observe the energy
density~(\ref{enden:dS}) produced by the de Sitter background.

\subsection{The General Case}
Now, we allow for a general expanding FLRW background given by the scale factor
$a(\eta)$, as well as for the scalar field $\phi$ a curvature coupling $\xi$
and a constant mass $m$.
Adiabatic expansion gives up to second order in
$d/d\eta$~\cite{GarbrechtProkopec:2004:2}
(here we keep the scale factor $a$ explicitly)
\begin{equation}
|\varphi|^2=\frac{1}{2\omega}-\frac{1}{4\omega^3}
\left\{
(6\xi-1)\frac{a^{\prime\prime}}{a}
-\frac 12 \frac{m^2 (aa^{\prime\prime}+{a^\prime}^2)}{\omega^2}
+\frac 54 \frac{m^4 a^2 {a^\prime}^2}{\omega^4}
\right\}
\end{equation}
and
\begin{eqnarray}
\Omega\!\!\!&=&\!\!\!
\omega+\frac{1}{2\omega}\left(\frac{a'^2}{a^2}+6\xi\frac{a''}{a}\right)
+\frac{1}{2}\frac{a'^2}{a^2}\frac{a^2m^2}{\omega^3}
+\frac18\frac{a'^2}{a^2}\frac{a^4m^4}{\omega^5}
\,,
\label{Omega:adiabatic}
\\
\label{Lambda:adiabatic}
\Lambda\!\!\!&=&\!\!\!
 \left\{
\frac{1}{2\omega}\left(\frac{a'^2}{a^2}\!+\!\frac{a''}{a}\right)
+\frac{1}{4}\bigg(
                  \frac{a''}{a}\!+\!3\frac{a'^2}{a^2}
             \bigg)\frac{a^2 m^2}{\omega^3}
-\frac 12 \frac{a'^2}{a^2}\frac{a^4 m^4}{\omega^5}
\right.
\\
&&\left.
+{\rm i}\frac{a'}{a}\bigg(
                    1+\frac12\frac{a^2m^2}{\omega^2}
              \bigg)
 \right\} {\rm e}^{-2{\rm i}\int^\eta \! W(\eta^\prime) d\eta^\prime}
\,,
\nonumber
\end{eqnarray}
where $\varphi(\eta)=(2W(\eta))^{-1/2}
\exp(-{\rm i}\int^\eta W(\eta^\prime)d\eta^\prime)$.
We therefore define
\begin{eqnarray}
\Delta_{A^2}\!\!\!&=&\!\!\!
\frac 1\omega\left\{
\frac{1-6\xi}{2}\frac{a^{\prime\prime}}{a}
+\frac 14 \frac{m^2 (aa^{\prime\prime}+{a^\prime}^2)}{\omega^2}
-\frac 58 \frac{m^4 a^2 {a^\prime}^2}{\omega^4}
\right\}\,,\\
\Delta_{\Omega}\!\!\!&=&\!\!\!\frac{1}{2\omega}\left(\frac{a'^2}{a^2}+6\xi\frac{a''}{a}\right)
+\frac{1}{2}\frac{a'^2}{a^2}\frac{a^2m^2}{\omega^3}
+\frac18\frac{a'^2}{a^2}\frac{a^4m^4}{\omega^5}
\,,
\end{eqnarray}
such that $|\varphi|^2=1/(2\omega)+\Delta_{A^2}/(2\omega^2)$
and $\Omega=\omega+\Delta_{\Omega}$.

The Lamb shift in FLRW Universe with respect to flat space is then
\begin{eqnarray}
\label{LambShift:FLRW}
\delta E\!\!&=&\!\!\delta E_{\rm FLRW}-\delta E_{\rm M}=
h^2 \int \frac{d^3k}{(2\pi)^3}
\left\{
\frac{1}{2\omega}
\frac{\Delta_{A^2}/\omega+1}{\Delta E -\omega-\Delta_\Omega}
-\frac{1}{2\omega}
\frac{1}{\Delta E - \omega}
\right\}\\
\!\!\!&\approx&\!\!\!\frac{h^2}{4\pi^2}
\int\limits_0^{\infty}dk \frac{k^2}{\omega^2}
\left\{
\frac{\Delta_{A^2}}{\Delta E - \omega}
+\frac{\omega\Delta_\Omega}{(\Delta E - \omega)^2}
\right\}
\nonumber\\
\!\!&=&\!\!\frac{h^2}{8\pi^2}\frac{1}{\Delta E}
\left\{
-\frac{5}{6}\frac{a^{\prime\prime}}{a}
- \frac{{a^\prime}^2}{a^2}
+(1-6\xi)
\log\Big(\frac{2\Delta E}{am}\Big)\frac{a^{\prime\prime}}{a}
+O\left(\frac{m a^{\prime\prime}}{\Delta E^2},
\frac{m {a^\prime}^2}{a\Delta E^2}\right)
\right\}\,,
\nonumber
\end{eqnarray}
where the relevant integrals are given in the appendix.
For $m \rightarrow 0$, there occurs a logarithmic infrared
divergence. This is however an artefact of adiabatic expansion, 
which breaks down in this limit.
Note in particular, that the exact
expression~(\ref{LambShift:deSitter})
for the massless de Sitter case is infrared finite.

Let us comment in more detail on the sensitivity of the detector to the energy
density produced by the expanding background. From the
expression~(\ref{LambShift}) we immediately see the contribution of
the mode energies $\Omega(\mathbf{k},\eta)$ to the Lamb shift, but additionally
there also enters the mode amplitude through
$\left|\varphi(\mathbf{k},\eta)\right|^2$, which is also 
influenced by the background. For this quantity, the relation 
\begin{equation}
\left|\varphi(\mathbf{k},\eta)\right|^2=\frac 12
\left(\Omega(\mathbf{k},\eta)-
\Re\left[\Lambda(\mathbf{k},\eta)
{\rm e}^{2{\rm i}\int^\eta W(\eta^\prime)d\eta^\prime}\right]\right)^{-1}
\end{equation}
generally holds. We hence found that
although we cannot write the result for the Lamb shift solely in terms of
$\Omega(\mathbf{k},\eta)$ or $T^0_{\phantom{0}0}$, we can  express it in
terms of contributions to the canonical Hamiltonian~(\ref{Hamiltonian:bosons}).

\section{Lamb Shift in Rindler Space}
It was suggested by Unruh~\cite{Unruh:1976}, that an accelerated
observer should perceive particles even in the vacuum, which is due to the
fact that quantization in a coordinate system suitable for the 
observer, referred to as Rindler space,
is inequivalent to quantization
in Minkowski space.
Therefore, accelerated observer vacuum and inertial Minkowski 
vacuum do not coincide~\cite{Fulling:1972}.
The quantum state in the accelerated system, which is equivalent to the
Minkowski vacuum, can be constructed through a Bogolyubov transformation,
which corresponds to mode mixing
and is known as the Unruh effect~\cite{Unruh:1976}.

Just as in de Sitter space, the response function of Unruh's detector falls off
exponentially~\cite{Unruh:1976,BroutMassarParentaniSpindel:1995}, therefore
resembling to a thermal spectrum. As we have observed for expanding Universes,
this effect is quantitatively dominated by the Lamb shift of energy levels.
In the following, we shall demonstrate that the same holds also true for an
accelerated detector.

\subsection{Scalar Field in Rindler Coordinates}
In flat two-dimensional space with the line element
\begin{equation}
  ds^2\equiv g_{\mu\nu}dx^\mu dx^{\nu}
         = - dt^2 + dx^2,\quad g_{\mu\nu}={\rm diag}(-1,1)\,,
\end{equation}
we consider an observer of mass $m_{\rm O}$, who is constantly accelerated
by the force $\mathbf{f}$, for example an ion in a homogeneous electric field.
Let us determine his trajectory $y(\tau)=(t(\tau),x(\tau))^T$, where
$\tau$ is his proper time, defined by $d\tau^2=-ds^2$.

The Minkowski vector describing the
force in the inertial system where the observer is instantaneously at rest is
\begin{equation}
\tilde f=
\left(
\begin{array}{c}
0\\
\mathbf{f}
\end{array}
\right)\,.
\end{equation}
When we see the observer moving at the instantaneous velocity $v$,
the force vector $f$ in our coordinate system is obtained from
\begin{equation}
f=\Lambda(-v) \tilde{f}=
\mathbf{f}\left(
\begin{array}{c}
\frac{v}{\sqrt{1-v^2}}\\
\frac{1}{\sqrt{1-v^2}}
\end{array}
\right)=
\mathbf{f}
\left(
\begin{array}{c}
\sinh \psi\\
\cosh \psi
\end{array}
\right)
\,,
\end{equation}
where $\Lambda(-v)$ denotes the Lorentz boost transformation, $\psi$
is the rapidity parameter, $\tanh \psi=v$,
and the velocity vector is of the standard form
\begin{equation}
u=\frac{dy}{d\tau}=
\left(
\begin{array}{c}
\frac{1}{\sqrt{1-v^2}}\\
\frac{v}{\sqrt{1-v^2}}
\end{array}
\right)
=
\left(
\begin{array}{c}
\cosh \psi\\
\sinh \psi
\end{array}
\right)
\,.
\end{equation}

With $p$ being his momentum,
the observer follows then a trajectory  which is
solution to the relativistic equation of motion
\begin{equation}
\frac{dp}{d\tau}=m_{\rm O}\frac{d^2y}{d\tau^2}=f\,.
\end{equation}
A solution for $dy/d\tau$ is easily found when setting $\psi=\alpha \tau$ and
$\alpha=\mathbf{f}/m_{\rm O}$, and we can interpret the parameter
$\alpha$ as a constant proper acceleration
\begin{equation}
\alpha=
\left[\left(\frac{d^2y}{d\tau^2}\right)^2\right]^\frac 12
=
\Big[{-\Big(\frac{d^2t}{d\tau^2}\Big)^2+
            \Big(\frac{d^2x}{d\tau^2}\Big)^2}\,\Big]^{\frac 12}
\,.
\end{equation}
A special $y(\tau)$ is given by
\begin{equation}
y(\tau)=
\left(
\begin{array}{c}
\alpha^{-1}\sinh \alpha\tau\\
\alpha^{-1}\cosh \alpha\tau
\end{array}
\right)
\,,
\end{equation}
implying the trajectory
\begin{equation}
x(t)=({t^2+\alpha^{-2}})^{1/2}\,,
\end{equation}
on which we shall consider Unruh's detector in the following.

Since we describe the time evolution of the detector in terms of its proper
time $\tau$, we also use $\tau$ as the time-variable for canonical
quantization of the scalar field, which then manifestly separates into modes
which the
observer perceives as of positive and of negative frequency, respectively. 
Let us therefore transform the system to the
Rindler coordinates as~\cite{Rindler:1966}
\begin{eqnarray}
\label{flat-Rindler:t}
t\!\!\!&=&\!\!\!\alpha^{-1} {\rm e}^\xi \sinh\alpha\tau
\,,
\\
x\!\!\!&=&\!\!\!\alpha^{-1} {\rm e}^\xi \cosh\alpha\tau\,,
\nonumber
\end{eqnarray}
such that the metric becomes
\begin{equation}
 ds^2=-{\rm e}^{2\xi}d\tau^2 + \alpha^{-2} {\rm e}^{2\xi}d\xi^2
\,,
\label{Rindler-metric}
\end{equation}
where the detector's site is at $\xi=0$.
The dependence of the metric~(\ref{Rindler-metric}) on $\xi$ indicates
that Minkowski space appears inhomogeneous to an accelerated observer.

According to the Lagrangean
\begin{equation}
\sqrt{-g}\mathcal{L}=\sqrt{-g}
\left(
-\frac 12 g_{\mu\nu}\partial^\mu\varphi\partial^\nu\varphi
-\frac 12 m^2 \varphi^2
\right)
\label{Lagrangean}
\,,
\end{equation}
the Klein-Gordon equation for a scalar field with mass $m$ is
\begin{equation}
\Big(-\frac{\partial^2}{\partial t^2}+\frac{\partial^2}{\partial x^2}
-m^2\Big)
\varphi(x,t)=0\,.
\end{equation}
In the Rindler coordinate system, this transforms to
\begin{equation}
\Big(-\frac{\partial^2}{\partial \tau^2}
+\alpha^2\frac{\partial^2}{\partial \xi^2}-{\rm e}^{2\xi}m^2
\Big)
\varphi(\xi,\tau)=0\,.
\label{Klein-Gordon:Rindler}
\end{equation}

We shall take the scalar field to be in the Rindler
vacuum $|0\rangle$, with the field operator 
\begin{equation}
\hat\varphi(\xi,\tau)=\int_{0}^{\infty}\frac{d\lambda}{2\pi}
                       \left\{c_\lambda \varphi_\lambda(\xi,\tau)
                      + c_\lambda^\dagger\varphi^*_\lambda(\xi,\tau)\right\}
\,,
\label{field-operator}
\end{equation}
where the creation and annihilation operators act as
$c_\lambda^\dagger|0\rangle=|\lambda\rangle$ and
$c_\lambda|\lambda^\prime\rangle=2\pi\delta(\lambda-\lambda^\prime)|0\rangle$.

Making use of the Jacobian of the coordinate 
transformation~(\ref{flat-Rindler:t}), 
$J = \alpha^{-1}{\rm exp}({2\xi})$, and the identity
$\int dt \int dx {\mathcal H}^\varphi (x,t)=
\int d\tau \int d\xi {\mathcal H}^\varphi(\xi,\tau)$,
one arrives at the following Hamiltonian
\begin{eqnarray}
\label{Hamiltonian:Rindler}
H^\varphi\!\!\!&=&\!\!\!\int_{-\infty}^{\infty} d\xi\, \mathcal{H}^\varphi
\,,
\\
\mathcal{H}^\varphi
\!\!\!&=&\!\!\! \frac{1}{2\alpha} \Big\{
\Big(\frac{\partial \hat\varphi}{\partial\tau}\Big)^2
+\alpha^2\Big(\frac{\partial \hat\varphi}{\partial\xi}\Big)^2
+{\rm e}^{2\xi}m^2\hat\varphi^2
\Big\}
\nonumber
\,.
\end{eqnarray}

The normalisable negative-frequency mode functions in~(\ref{field-operator})
can be expressed in terms of Hankel functions as
({\it cf. } Refs.~\cite{Fulling:1972,Boulware:1974,Gerlach:1989})
\begin{equation}
\varphi_\lambda(\xi,\tau)=
{\rm e}^{-{\rm i}\lambda\tau}{\rm e}^{-\frac{\pi}{2} \frac{\lambda}{\alpha}}
\sqrt{\frac{\pi}{2\alpha} \sinh\left(\pi\frac{\lambda}{\alpha}\right)}
\,
{\rm i}H^{(1)}_{{\rm i}\frac{\lambda}{\alpha}}
 \left({\rm i}\frac{m}{\alpha}{\rm e}^\xi \right)
.
\label{Rindler Modes}
\end{equation}
%
In order to fix the normalisation, we define the
scalar product as implied by Green's theorem~\cite{HawkingEllis:1973}
\begin{equation}
(\varphi_\lambda,\varphi_\lambda^\prime)
={\rm i}\int\limits_\Sigma d \Sigma^\mu \sqrt{-g}
\left(\varphi_\lambda^*  \stackrel{\leftrightarrow}{\partial_\mu}
      \varphi_{\lambda^\prime}\right)\,,
\end{equation}
where $\Sigma$ is a \emph{spatial} hypersurface, for example the surface
defined by $\tau=0$, which we consider in the following.
When intoducing
\begin{equation}
\nonumber
{\cal N}_\lambda
=
{\rm e}^{-\frac{\pi}{2}\frac{\lambda}{\alpha}}
\sqrt{\pi\frac{\lambda}{\alpha}
\sinh\left(\pi\frac{\lambda}{\alpha}\right)
}\,,
\end{equation}
$\varrho=\alpha^{-1} {\rm e}^\xi$ and an infinitesimal regulator
$\varepsilon$,
the scalar product turns out to be the desired $\delta$-function
representation
\begin{eqnarray}
\label{KG:ScalarProduct}
(\varphi_\lambda,\varphi_\lambda^\prime)
\!\!\!&=&\!\!\!{-\rm i}\int\limits_{-\infty}^{\infty} d\xi \frac{1}{\alpha}
\left(\varphi_\lambda^*  \stackrel{\leftrightarrow}{\partial_\tau}
      \varphi_{\lambda^\prime}\right)
\\
\nonumber
\!\!\!&=&\!\!\!
{\cal N}_\lambda {\cal N}_{\lambda^\prime}
\int\limits_{0}^{\infty} \frac{d\varrho}{\alpha\varrho^{1+\varepsilon}}
{H^{(1)}_{{\rm i}\frac{\lambda}{\alpha}}}^*
\left({\rm i} m \varrho\right)
H^{(1)}_{{\rm i}\frac{\lambda^\prime}{\alpha}}
\left({\rm i} m \varrho\right)
\\
\nonumber
\!\!\!&=&\!\!\!
{\cal N}_\lambda {\cal N}_{\lambda^\prime}
\frac{1}{8\alpha}
\frac{\left|
\Gamma\left(\varepsilon+{\rm i}\frac{\lambda+\lambda^\prime}{2\alpha}\right)
\right|^2
+\left|
\Gamma\left(\varepsilon+{\rm i}\frac{\lambda-\lambda^\prime}{2\alpha}\right)
\right|^2
}
{\Gamma(\varepsilon)}\\
\nonumber
\!\!\!&\approx&\!\!\!
{\cal N}_\lambda {\cal N}_{\lambda^\prime}
\frac{1}{2\alpha}
\left|
\Gamma\left(\frac{\lambda+\lambda^\prime}{2\alpha}\right)
\right|^2
\frac{\varepsilon}{\varepsilon^2+
\left(\frac{\lambda-\lambda^\prime}{\alpha}\right)^2}
=2\pi\delta(\lambda-\lambda^\prime)
\,.
\end{eqnarray}
For solving the integral, we made use of the formula GR~6.576.4
(we denote the equalities taken from Ryzhik and
Gradshteyn~\cite{GradshteynRyzhik:1965} by GR).

It is also of interest to compute the Bogolyubov coefficients for the
matching to the Minkowski modes
\begin{equation}
\psi_k(x,t)=\frac{1}{\sqrt{2\omega}}{\rm e}^{-{\rm i}\omega t + {\rm i} k x}
\,,
\end{equation}
where $\omega=\sqrt{k^2+m^2}$.
When we use the formula GR~6.621.3 for the occuring integrals and
GR~9.121.19, GR~9.121.21 in order to express hypergeometric functions 
in terms of elementary functions, we obtain (\emph{cf.}~\cite{Fulling:1972})
\begin{eqnarray}
\alpha_{\lambda,k}\!\!\!&=&\!\!\!(\psi_k,\phi_\lambda)
={\rm e}^{-\frac{\pi}{2}\frac{\lambda}{\alpha}}
\sqrt{\frac{\pi}{2\alpha\omega}\sinh\left(\pi \frac{\lambda}{\alpha}\right)}
\int\limits_0^\infty d\varrho 
\left(\omega +\frac{\lambda}{\alpha\varrho}\right)
{\rm e}^{-{\rm i}k \varrho}
H^{(1)}_{{\rm i}\frac{\lambda}{\alpha}}({\rm i} m \varrho)
\\
\nonumber
\!\!\!&=&\!\!\!
(2m)^{{\rm i}\frac{\lambda}{\alpha}}
\sqrt{\frac{\pi}{\omega\lambda}}
\left(1-{\rm e}^{-2 \pi\frac{\lambda}{\alpha}}\right)^{-\frac 12}
\left(\frac{\omega+k}{m}\right)^{{\rm i}\frac{\lambda}{\alpha}}
\,,
\end{eqnarray}
and likewise
\begin{equation}
\beta_{\lambda,k}=(\psi_k^*,\phi_\lambda)
=
(2m)^{{\rm i}\frac{\lambda}{\alpha}}
\sqrt{\frac{\pi}{\omega\lambda}}
\left({\rm e}^{2 \pi\frac{\lambda}{\alpha}}-1\right)^{-\frac 12}
\left(\frac{\omega-k}{m}\right)^{-{\rm i}\frac{\lambda}{\alpha}}
\,.
\end{equation}
The $k$-dependent phase is missed when matching
instead of along $\tau=t=0$ along
$U=t-x$ and $V=t+x$~\cite{Unruh:1976}, because the Rindler modes
reduce to massless modes along these lightlike surfaces and Green's theorem
is strictly speaking only applicable along proper spacelike surfaces.

Note that the mode mixing is exponentially suppressed, since
 $\beta_{\lambda,k}
 \propto {\rm e}^{-\pi \lambda/\alpha}$
for large $\lambda$.
Assuming $m\gg \alpha$, as required by the expansions we shall use in the
following,  $\beta_{\lambda,k}\ll 1$, and
one does not need to account for the mode mixing. In order not
to distract from the main line of argument, we therefore neglect it here.

The spatial parts of the Rindler modes~(\ref{Rindler Modes}) are real valued
functions of $\xi$, because the operator~(\ref{Klein-Gordon:Rindler})
corresponds to a quantum mechanical particle which is reflected by
a potential ${\rm e}^{2\xi} m$~\cite{Fulling:1972}.
For $\xi \rightarrow -\infty$ the
solutions therefore reduce to standing plane waves.
This picture also explains the exponential decay of the mode functions
for $m\varrho\gg \lambda/\alpha$, which is by GR 8.451.3
\begin{equation}
H^{(1)}_{{\rm i}\frac{\lambda}{\alpha}}({\rm i}m\varrho)
\sim -{\rm i}\sqrt{\frac{2}{\pi m\varrho}}
{\rm e}^{-m{\varrho}+\frac{\pi}{2}\frac{\lambda}{\alpha}}
\,.
\end{equation}
As a consequence, the condition $\lambda \gg \alpha$ effectively holds
whenever $m \gg \alpha$, due to the exponential vanishing of $\varphi$
at the detector's site $\varrho=\alpha^{-1}$ when $\lambda < m$.
For calculational
purposes and in order to keep notations simple, it proves useful to
complexify the expansion. First, note the identity~\cite{Watson:1966}
(we introduce an infinitesimal real part $\varepsilon>0$ of the argument
of the Bessel function for later use)
\begin{eqnarray}
\label{useful-relations}
 J_{{\rm i}\lambda/\alpha}\Big({\rm i}\frac{m}{\alpha}{\rm e}^{\xi}-\varepsilon\Big)
    \!\!\!&=&\!\!\! \Big[
           J_{-{\rm i}\lambda/\alpha}
       \Big({\rm e}^{-{\rm i}\pi}{\rm i}\frac{m}{\alpha}{\rm e}^{\xi}
                                   +{\rm e}^{-{\rm i}\pi}\varepsilon\Big)
      \Big]^*
\\
    \!\!\!&=&\!\!\! {\rm e}^{-\frac{\pi\lambda}{\alpha}}
       \Big[
           J_{-{\rm i}\lambda/\alpha}
     \Big({\rm i}\frac{m}{\alpha}{\rm e}^{\xi}+\varepsilon\Big)
      \Big]^*
\,,
\nonumber
\end{eqnarray}
which, together with the definitions GR~8.403.1 and GR~8.405.1,
allows us to express
\begin{equation}
H^{(1)}_{{\rm i}\frac{\lambda}{\alpha}}
\left({\rm i}\frac{m}{\alpha}{\rm e}^\xi\right)=
\frac{4}{1-{\rm e}^{-2\pi \frac{\lambda}{\alpha}}}
\Im \left[J_{{\rm i}\frac{\lambda}{\alpha}}
\left({\rm i}\frac{m}{\alpha}{\rm e}^\xi\right)\right]
\,,
\end{equation}
and therefore
\begin{equation}
\varphi_\lambda(\xi,\tau)
=\frac{{\rm e}^{-{\rm i}\lambda\tau +\frac{\pi}{2}\frac{\lambda}{\alpha}}}
      {\sqrt{2\lambda}}
 \left|\Gamma\left(1+{\rm i}\frac{\lambda}{\alpha}\right)\right|
\left\{ J_{{\rm i}\frac{\lambda}{\alpha}}
\left({\rm i}\frac{m}{\alpha}{\rm e}^\xi + \varepsilon\right)
-{\rm e}^{-\pi\frac{\lambda}{\alpha}}
J_{-{\rm i}\frac{\lambda}{\alpha}}
\left({\rm i}\frac{m}{\alpha}{\rm e}^\xi -\varepsilon \right)
\right\}
\,,
\end{equation}
where $\lambda>0$.
We now allow for $\lambda\in[-\infty,\infty]$, introduce the
complex mode functions
\begin{equation}
\tilde\varphi_\lambda(\xi,\tau)=
\frac{{\rm e}^{-{\rm i}|\lambda|\tau}}{\sqrt{2|\lambda|}}
\,{\rm e}^{\frac{\pi}{2}\frac{\lambda}{\alpha}}
\left| \Gamma\Big(1\!+\!{\rm i}\frac{\lambda}{\alpha}\Big) \right|
J_{{\rm i}\frac{\lambda}{\alpha}}
 \left({\rm i}\frac{m}{\alpha}{\rm e}^\xi+{\rm sign}(\lambda)\varepsilon\right)
\label{Rindler Modes Complex}
\,,
\end{equation}
and reexpress the field operator~(\ref{field-operator}) as
\begin{equation}
\hat\varphi(\xi,\tau)=\int_{-\infty}^{\infty}\frac{d\lambda}{2\pi}
                       \left\{\tilde c_\lambda \tilde\varphi_\lambda(\xi,\tau)
                      + \tilde c_\lambda^\dagger \tilde\varphi^*_\lambda(\xi,\tau)\right\}
\,,
\label{field-operator:complex}
\end{equation}
where the states are restricted to those generated by the pairs
$\tilde c_\lambda^\dagger + \tilde c_{-\lambda}^\dagger$ acting on
the ground state $|0\rangle$.

When the parameter $\lambda$ becomes large compared to
the acceleration $\alpha$, the
modes~(\ref{Rindler Modes},~\ref{Rindler Modes Complex}) asymptotically
reduce to plane waves.
We provide here a systematic expansion of the mode
functions~(\ref{Rindler Modes Complex}) in Rindler space in the ultraviolet domain,
that is where $\lambda\gg\alpha,m$.  All terms involving powers
up to $\alpha^2$ and $m^4$ are displayed.
Since $|\lambda|/\alpha\gg 1$, we need an asymptotic expansion of
Bessel functions of large order, which is given by the 
approximation by tangents~\cite{Watson:1966}:
\begin{eqnarray}
\label{Bessel:Approx:Positive}
J_{{\rm i}\frac{\lambda}{\alpha}}
\Big({\rm i} \frac{m}{\alpha}{\rm e}^\xi+\varepsilon\Big)
\!\!\!&\sim&\!\!\!
\frac{{\rm{e}^{{\rm i}\frac{\lambda}{\alpha}\left(\tanh \beta-\beta\right)
               -\frac 14 \pi {\rm i}}}}
     {\big(2\pi\frac{\lambda}{\alpha} \tanh \beta\big)^{1/2}}
 \left\{
1-{\rm i} \frac{\alpha}{\lambda}\left(
\frac 18 \coth \beta - \frac 5{24} \coth^3 \beta
\right)
\right.
\\
&&
\hskip -2.5cm
\left.
-\frac{\alpha^2}{\lambda^2}
\left(
\frac{9}{128}\coth^2 \beta -\frac{231}{576}\coth^4 \beta
+\frac{1155}{3456} \coth^6 \beta
\right)
+O\left(\frac{\alpha^3}{\lambda^3}\right)
\right\},\, {\rm for}\,\lambda>0
,\quad
\nonumber
\end{eqnarray}
where 
\begin{eqnarray}
  \cosh \beta = \frac{\lambda}{m}{\rm e}^{\xi} \!+\! {\rm i}\varepsilon
,\;
\;\;
\tanh \beta
             = \Big(1-\frac{m^2}{\lambda^2}{\rm e}^{2\xi}\Big)^{\frac 12}
.\nonumber
\end{eqnarray}
The expansion~(\ref{Bessel:Approx:Positive}) corresponds to the region 
2 of figure 22 in Ref.~\cite{Watson:1966}.

When $\lambda < 0$, we make use of the identity~(\ref{useful-relations})
to bring the argument of the Bessel function in~(\ref{Rindler Modes Complex})
into the region of validity ($|{\rm arg}(z)|<\pi/2$)
of the approximation by tangents. 
Indeed, since the argument fulfills 
${\rm arg}\big({\rm i}({m}/{\alpha}){\rm e}^{\xi}+\varepsilon\big)<\pi/2$,
we can use the approximation by tangents
($\cosh(\beta)=\nu/z$ 
lies again in the region 2 of figure 22 in Ref.~\cite{Watson:1966})
\begin{eqnarray}
\label{Bessel:Approx:Negative}
J_{{\rm i}\frac{\lambda}{\alpha}}
   \Big({\rm i} \frac{m}{\alpha}{\rm e}^\xi-\varepsilon\Big)
\sim {\rm e}^{-\pi \lambda/\alpha}
\frac{{\rm e}^{{\rm i}\frac{\lambda}{\alpha}\left(\tanh \beta-\beta\right)
               +\frac 14 \pi i}}
     {\big(-2\pi\frac{\lambda}{\alpha} \tanh \beta\big)^{1/2}}
 \left\{
1-{\rm i} \frac{\alpha}{\lambda}\left(
\frac 18 \coth \beta - \frac 5{24} \coth^3 \beta
\right)
\right.
\\
\left.
-\frac{\alpha^2}{\lambda^2}
\left(
\frac{9}{128}\coth^2 \beta -\frac{231}{576}\coth^4 \beta
+\frac{1155}{3456} \coth^6 \beta
\right)
+O\left(\frac{\alpha^3}{\lambda^3}\right)
\right\}\,, {\rm for}\,\lambda<0
.\quad
\nonumber
\end{eqnarray}
The following expressions, which are valid for both for $\lambda<0$
and $\lambda>0$, completely specify $\beta$, 
\begin{eqnarray}
\coth \beta \!\!\!&=&\!\!\! \Big(1-\frac{m^2}{\lambda^2} {\rm e}^{2\xi}\Big)^{-\frac 12}
\,,
\label{coth}
\\
  \cosh \beta \!\!\!&=&\!\!\! \frac{|\lambda|}{m}{\rm e}^{\xi} 
                  - {\rm i}\varepsilon{\rm sign}(\lambda)
\,.
\label{cosh:2}
\end{eqnarray}

We can now write a general approximation by tangents for the Rindler
modes~(\ref{Rindler Modes}) (valid for any $|\lambda|\gg \alpha, m$):
\begin{eqnarray}
\tilde\varphi_\lambda \!\!\! &\sim&\!\!\! 
  \frac{{\rm e}^{-{\rm i}|\lambda|\tau +\frac{\pi}{2}\frac{|\lambda|}{\alpha}}}
       {\sqrt{2|\lambda|}}
\left|\Gamma\Big(1\!+\!{\rm i}\frac{\lambda}{\alpha}\Big)\right|
\frac{{\rm e}^{{\rm i}\frac{\lambda}{\alpha}\left(\tanh \beta-\beta\right)
               -{\rm i}\frac{\pi}{4}{\rm sign}(\lambda)}}
     {\big[2\pi(|\lambda|/\alpha) \tanh \beta\big]^{1/2}}
\left\{
1\!-\!{\rm i} \frac{\alpha}{\lambda}\Big(
\frac 18 \coth \beta - \frac 5{24} \coth^3 \beta
\Big)
\right.
\nonumber
\\
&&
\left.
-\frac{\alpha^2}{\lambda^2}
\left(
\frac{9}{128}\coth^2 \beta -\frac{231}{576}\coth^4 \beta
+\frac{1155}{3456} \coth^6 \beta
\right)
+O\left(\frac{\alpha^3}{\lambda^3}\right)
\right\}
\label{Rindler Modes:Tangents}
\end{eqnarray}

Upon expanding $\varphi_\lambda$ in powers of $\alpha/\lambda$,
$(m/\lambda)^2$ and $\xi$ we get (up to corrections of order 
$O\big((\alpha/\lambda)^3,\xi^2,(m/\lambda)^6\big)$,
\begin{eqnarray}
\label{RindlerModes:Expanded}
\tilde\varphi_\lambda \!\!\!&\simeq&\!\!\!
                        {\rm e}^{-{\rm i}|\lambda|\tau}
                        \frac{\alpha^{1/2}}{(4\pi)^{1/2}|\lambda|}
                        {\rm e}^{\frac{\pi}{2}\frac{|\lambda|}{\alpha}
                                 - {\rm i}\frac{\pi}{4}{\rm sign}(\lambda)}
                   \left|\Gamma\Big(1+{\rm i}\frac{\lambda}{\alpha}\Big)\right|
\\
&&\!\!\!\times
                {\rm exp}\left[{\rm i}\frac{\lambda}{\alpha}
                                \left(
                                     1-\log\left(\frac{2|\lambda|}{m}\right)
                                  -  \frac 14\frac{m^2}{\lambda^2}
                                  -  \frac 18\frac{m^4}{\lambda^4}
                                \right)\right]
                    \times
                     {\rm exp}\left[{\rm i}\xi\frac{\lambda}{\alpha}
                              \left(1\!-\!\frac{m^2}{2\lambda^2}\!-
                              \!\frac{m^4}{8\lambda^4}\right)
                          \right]
\nonumber
\\
&&\!\!\!\times
\left\{
1+\frac 14 \frac{m^2}{\lambda^2} +\frac{5}{32}\frac{m^4}{\lambda^4}
+\frac 12 \xi \frac{m^2}{\lambda^2} +\frac 58 \xi \frac{m^4}{\lambda^4}
\right\}
\nonumber
\\
&&\!\!\!\times
\left\{ 1+ \frac{\rm i}{12}\frac{\alpha}{\lambda}
           \left[1+3\frac{m^2}{\lambda^2}+\frac{33}{8}\frac{m^4}{\lambda^4}
             + 6 \xi \frac{m^2}{\lambda^2}+\frac{33}{2}\xi\frac{m^4}{\lambda^4}
           \right]\right.\nonumber\\
          &&\;\;\;\,\left.
          -\frac{1}{288}\frac{\alpha^2}{\lambda^2}
          \left[1+78\frac{m^2}{\lambda^2}+\frac{1005}{4}\frac{m^4}{\lambda^4}
             + 156 \xi \frac{m^2}{\lambda^2}+1005\xi\frac{m^4}{\lambda^4}
          \right]
     \right\}\,.
\nonumber
\end{eqnarray}
For $\lambda\rightarrow\infty$, this reduces to a plane wave solution as
it should, since the acceleration parameter $\alpha$ becomes irrelevant.
Investigating the $\xi$-dependent part of this expansion, we
note
\begin{equation}
\exp\left[{\rm i}\xi\frac{\lambda}{\alpha}
                                \left(1
                                   - \frac{m^2}{2\lambda^2}
                                  -  \frac{m^4}{8\lambda^4}
                                  +O\left(\frac{m^6}{\lambda^6}\right)
                                \right)\right]
=
\exp\left[
{\rm i}\xi \frac{1}{\alpha}\sqrt{\lambda^2-m^2}
\right]
\,.
\label{klambda}
\end{equation}
Therefore, we can interpret the modes locally as particles of energy
$\lambda$ and momentum $\sqrt{\lambda^2-m^2}$.

\subsection{Lamb Shift}
In order to calculate the Lamb shift, we need the amplitude squared of the
expanded mode functions~(\ref{RindlerModes:Expanded}), which is at the site
$\xi=0$
\begin{equation}
|\tilde\varphi_\lambda(\xi=0,\tau)|^2=\frac{1}{2\lambda}
\frac{1}{1-{\rm e}^{-2\pi|\lambda|/\alpha}}
\left(1+\frac{1}{2}\frac{m^2}{\lambda^2}
+\frac{3}{8}\frac{m^4}{\lambda^4}+\ldots\right)
\left(1+\frac 12 \frac{\alpha^2 m^2}{\lambda^4}+\ldots\right)
\,.
\label{phi:modsquare}
\end{equation}

The Hamiltonian~(\ref{Hamiltonian:Rindler}), which is quadratic in
the field operators~(\ref{field-operator}),
can be recast into the following 
quadratic form in terms of the creation and annihilation
operators of the Rindler vacuum, 
\begin{eqnarray}
H^\varphi =\frac 12 \!\int\frac{d\lambda}{2\pi}\!
\big\{
 \Omega_\lambda
 \big(\tilde c_\lambda \tilde c^\dagger_\lambda
      +\tilde c^\dagger_\lambda\tilde  c_\lambda \big)
 +\,\big(\Lambda_\lambda\tilde  c_\lambda \tilde c_{-\lambda}\!+\!{\rm h.c.}\big)
\big\}
\label{Hamiltonian:Generic}
\,,\quad
\end{eqnarray}
where $\Omega_\lambda$ denotes the
virtual energy of a Rindler quasiparticle excitation
of momentum $\lambda$, and $\Lambda_\lambda$ is the amplitude for 
annihilation of a Rindler pair, with the momenta $\lambda$ and $-\lambda$,
respectively.

Here, we are interested in the ultraviolet domain, {\it i.e.} in
the portion of~(\ref{Hamiltonian:Generic}) where $|\lambda|\gg \alpha$.
Thus, it is possible to choose $1\gg\Delta\xi\gg \alpha/|\lambda|$, 
which is what we assume in the following. 
From the analytic behaviour of the Rindler modes~(\ref{Rindler Modes}),
it then follows that in the detector's neighbourhood at $\xi = 0$ ,
the ultraviolet contributions to the Hamiltonian~(\ref{Hamiltonian:Generic})
are dominated by
the local contribution from $\xi \in (-\Delta\xi/2, \Delta\xi/2)$,
\begin{eqnarray}
H^\varphi\approx \int_{-\Delta\xi/2}^{\Delta\xi/2}d\xi\, \mathcal{H}^\varphi
\label{Hamiltonian:Integral}
\,.
\end{eqnarray}
Then, by using the following approximate relation,
\begin{eqnarray}
\int_{-\Delta\xi/2}^{\Delta\xi/2}d\xi
          {\rm e}^{{\rm i}\xi\frac{\lambda}{\alpha}
             \big(1-\frac12\frac{m^2}{\lambda^2}
                     -\frac18\frac{m^4}{\lambda^4}  \big)
\pm
{\rm i}\xi\frac{\lambda^\prime}{\alpha}
             \big(1-\frac12\frac{m^2}{\lambda^{\prime 2}}
                     -\frac18\frac{m^4}{\lambda^{\prime 4}}  \big)
}
\!\!\!&\approx&\!\!\! 2\pi\alpha\delta(\lambda\pm \lambda^\prime)
                \left(1-\frac12\frac{m^2}{\lambda^2}
                       -\frac18\frac{m^4}{\lambda^4}\right)
\,,
\end{eqnarray}
we find
\begin{eqnarray}
\Omega_\lambda\!\!\!&=&\!\!\!\left\{
(\lambda^2+m^2)|\varphi_\lambda|^2+ \alpha^2|\partial_\xi\varphi_\lambda|^2
\right\}\left(1-\frac12\frac{m^2}{\lambda^2}
                       -\frac18\frac{m^4}{\lambda^4}\right)
\,,\\
\Lambda_\lambda\!\!\!&=&\!\!\!\left\{
(m^2-\lambda^2)|\varphi_\lambda|^2 + \alpha^2|\partial_\xi\varphi_\lambda|^2
\right\}\left(1-\frac12\frac{m^2}{\lambda^2}
                       -\frac18\frac{m^4}{\lambda^4}\right)
        {\rm e}^{-2{\rm i}|\lambda|\tau}
\,.
\end{eqnarray}
Assuming $m\gg \alpha$,
by the same token as for neglecting the mode mixing, we have dropped here the
exponentially falling prefactor occuring in expression~(\ref{phi:modsquare}).

Upon substituting the expanded mode functions~(\ref{RindlerModes:Expanded}),
one obtains after some algebra
\begin{eqnarray}
\Omega_\lambda\!\!\!&=&\!\!\!
\frac {1}{|\lambda|} \left(
\lambda^2  - \frac 38 \frac{\alpha^2 m^4}{\lambda^4}
+\ldots
\right)
\,,\\
\Lambda_\lambda\!\!\!&=&\!\!\! - \frac{\alpha^2m^2}{2|\lambda|^3}
                      \,{\rm e}^{-2i|\lambda|\tau}
+\ldots
\label{Lambda:Rindler}
\end{eqnarray}
%
%

We can now assemble the difference between Lamb shift in Rindler
space $\delta E_R$
and flat space $\delta E_M$. For the lower limit of the self-energy
integral we take $\lambda=m$, because for $\lambda<m$ the modes
are exponentially suppressed.
According to the two-dimensional case
of Eqn.~(\ref{LambShift}), we find
\begin{eqnarray}
\delta E\!\!&=&\!\!\delta E_R - \delta E_M
=
2 h^2 \int\limits_{m}^{\infty} \frac{d\lambda}{2\pi}\frac{1}{2\lambda}
\left(1+\frac 12 \frac{m^2}{\lambda^2}+\frac 38 \frac{m^4}{\lambda^4}\right)
\left\{
\frac{1+\frac{\alpha^2m^2}{2\lambda^4}}
{\Delta E-\lambda+\frac 38 \frac{\alpha^2m^4}{\lambda^5}}
-\frac{1}{\Delta E-\lambda}
\right\}
\label{Lamb:Rindler}
\\
\!\!\!&\approx&\!\!\!
\frac{h^2}{4\pi} \int\limits_{0}^{\infty}dk
 \frac{\alpha^2m^2}{\lambda^5}
\left\{
\frac{1}
{\Delta E-\lambda}
-\frac{3}{4\lambda}
\frac{m^2}
{(\Delta E-\lambda)^2}
\right\}\nonumber
= \frac{h^2}{6\pi}\frac{\alpha^2}{\Delta E m^2}\Big(1 + O(m/\Delta E)\Big)
\end{eqnarray}
where 
we have identified the momentum according to the identity~(\ref{klambda}) as
$k = \sqrt{\lambda^2-m^2}$ and substituted $d\lambda \rightarrow dk$. 
The integrals are evaluated according
to~(\ref{LambRindlerInt:Amp},~\ref{LambRindlerInt:Omega}), and the final result
is displayed up to leading order in $1/\Delta E$, assuming that
$\Delta E\gg m$. 
When compared with the exponentially falling particle number by mode mixing,
we see that in the ultraviolet, Unruh effect gets a boost.

\section{Lamb Shift Versus Response Rate}

While the response rate of an Unruh detector falls off exponentially with the
level spacing $\Delta E$, which holds true for de Sitter as well as for Rindler
space, we have shown that Lamb shift exhibits a power-law behaviour.
From the response function
$d{\cal F}(\Delta E)/d\tau$~(\ref{ResponeFunction:m=0}) for positive and
negative $\Delta E$, one can derive according to the principle of
detailed balance the probability to find the detector
on an excited level when being in equilibrium with the background,
see \emph{e.~g.} Ref.~\cite{GarbrechtProkopec:2004:1}. This probability
turns out to be exponentially falling with $\Delta E$ too.
Since also Lamb shift corresponds to the mixing of energy levels
both effects are therefore
quantitatively comparable, and Lamb shift is clearly more important in the
ultraviolet. Yet, the difference is of course that the response of the
detector is a time-dependent while Lamb shift a time-independent
effect\footnote{
Strictly speaking,
the Lamb shift in the expanding Universe~(\ref{LambShift:FLRW})
varies as the Hubble rate changes with time. Time-independence means here
that we use time-independent perturbation theory. 
}.

For the expanding Universe,
the power law behaviour is expected when considering the Hamiltonian
or the local covariant energy density. For the accelerated observer, we have
derived a new expression for a local
virtual energy density, which also
corresponds to a power law of the mode-energy.
From this point of
view, the exponential decay of the detector response comes out as a surprise.
Note, that for the calculation of Lamb shift,
we had to perform renormalisations, though at a rather crude technical level.
Therefore, we would have to verify whether the response
function~(\ref{ResponseFunction}) for the unrenormalised,
bare detector correctly reproduces the response rate for its
renormalised, dressed counterpart. In QFT, this question is answered positively
by the LSZ reduction formula for scattering amplitudes,
the proof of which in particular requires that the external
states of the matrix element correspond to well separated wave packets.
It is not clear whether this condition can be met for the state being the
product of detector in the ground state and curved spacetime vacuum. In
particular, our discussion of boundary effects in
Ref.~\cite{GarbrechtProkopec:2004:1} indicates possible problems,
since a huge period of interaction with the vacuum and a tremendous
coherence is required, while scattering in flat space is a resonance
phenomenon on rather short timescales. However, we do not
decide this question in this paper.

Since the Lamb shift for the Unruh detector can be interpreted as a
self-energy, it is interesting to compare with the effects of
self-energies for quantum fields. In
Refs.~\cite{ProkopecTornkvistWoodard:2002:1,ProkopecTornkvistWoodard:2002:2,ProkopecPuchwein:2003},
the vacuum polarization of a photon coupled to minimally
and nearly minimally coupled scalars in de Sitter space is calculated.
As a main result, a one-loop effective equation of motion is derived,
which indicates a mass term for the photon, but no damping term corresponding
to scattering from a thermal scalar background.

Therefore, self-energy effects known for fields in curved spacetimes
are also of relevance for the Unruh detector. In fact, it is not \emph{via}
the response rate~\footnote{
We were unaware of this when we wrote Ref.~\cite{GarbrechtProkopec:2004:2}.
}
but through the Lamb shift how a bound state probes
the quantum vacuum and the energy produced by the background.

\appendix*
\section{Integrals for Lamb Shift Calculation\label{App:Integrals}}

For the calculation of Lamb shift in FLRW-background, we need
the following integrals:
\begin{eqnarray}
I_1 \!\!\!&=&\!\!\! \int
\frac{k^2 dk}{(\Delta E -\omega)\omega^3}
\\
\!\!\!&=&\!\!\!
-\frac{k}{\Delta E \omega} -\frac{m}{\Delta E^2}\arctan \frac{k}{m}
\nonumber\\
&+&\frac{\sqrt{\Delta E^2-m^2}}{\Delta E^2}\frac 12
\Bigg\{
\log\left|
\frac{\sqrt{\Delta E^2-m^2}+k}{\sqrt{\Delta E^2-m^2}-k}
\right|
+
\log\left|
\frac{\sqrt{\Delta E^2-m^2}\omega+\Delta E k}
     {\sqrt{\Delta E^2-m^2}\omega-\Delta E k}
\right|
\Bigg\}
\nonumber
\,,
\\&&\phantom{Spacer} \nonumber\\
I_2 \!\!&=&\!\! \int 
\frac{k^2 dk}{(\Delta E -\omega)\omega^5}
\\
\!\!&=&\!\!
   \frac{1}{\Delta E}\frac{k^3}{3m^2\omega^3}
 + \frac{1}{\Delta E^2}\Bigg[
                            \frac{1}{2m}\arctan\Big(\frac{k}{m}\Big)
                          - \frac{k}{2\omega^2}
                       \Bigg]
 - \frac{1}{\Delta E^3}\frac{k}{\omega} 
 -  \frac{m}{\Delta E^4}\arctan\Big(\frac{k}{m}\Big)
\nonumber\\
&+&\frac{\sqrt{\Delta E^2-m^2}}{\Delta E^4}\frac 12
\Bigg\{
\log\left|
\frac{\sqrt{\Delta E^2-m^2}+k}{\sqrt{\Delta E^2-m^2}-k}
\right|
+
\log\left|
\frac{\sqrt{\Delta E^2-m^2}\omega+\Delta E k}
     {\sqrt{\Delta E^2-m^2}\omega-\Delta E k}
\right|
\Bigg\}
\nonumber
\,,
\\&&\phantom{Spacer} \nonumber\\
I_3 \!\!\!&=&\!\!\! \int 
\frac{k^2 dk}{(\Delta E -\omega)\omega^7}
\\
\!\!&=&\!\!
   \frac{1}{\Delta E}\Bigg[
                           \frac{2}{15}\frac{k}{m^4\omega}
                        +  \frac{1}{15}\frac{k}{m^2\omega^3}
                        -  \frac{1}{5}\frac{k}{\omega^5}
                     \Bigg]
  + \frac{1}{\Delta E^2}\Bigg[
                            \frac{1}{8m^3}\arctan\Big(\frac{k}{m}\Big)
                          + \frac{k}{8m^2\omega^2}
                          - \frac14\frac{k}{\omega^4}
                       \Bigg]
\nonumber\\
      &+&  \frac{1}{\Delta E^3}\frac13\frac{k^3}{m^2\omega^3} 
       +  \frac{1}{\Delta E^4}\Bigg[
                                    \frac{1}{2m}\arctan\Big(\frac{k}{m}\Big)
                                - \frac12 \frac{k}{\omega^2}
                              \Bigg]
       - \frac{1}{\Delta E^5}\frac{k}{\omega}
       -  \frac{m}{\Delta E^6}\arctan\Big(\frac{k}{m}\Big)
\nonumber\\
&+&\frac{\sqrt{\Delta E^2-m^2}}{\Delta E^6}\frac 12
\Bigg\{
\log\left|
\frac{\sqrt{\Delta E^2-m^2}+k}{\sqrt{\Delta E^2-m^2}-k}
\right|
 +
\log\left|
\frac{\sqrt{\Delta E^2-m^2}\omega+\Delta E k}
     {\sqrt{\Delta E^2-m^2}\omega-\Delta E k}
\right|
\Bigg\}
\nonumber
\,,
\end{eqnarray}

\begin{eqnarray}
J_1 \!\!\!&=&\!\!\! \int 
\frac{k^2 dk}{(\Delta E -\omega)^2\omega^2}
\\*
\!\!\!&=&\!\!\!
 \frac{k}{\Delta E(\Delta E - \omega)}
-\frac{m}{\Delta E^2}\arctan \frac{k}{m}
\nonumber\\
&-&\frac{m^2}{\Delta E^2\sqrt{\Delta E^2-m^2}}\frac 12
\Bigg\{
\log\left|
\frac{\sqrt{\Delta E^2-m^2}+k}{\sqrt{\Delta E^2-m^2}-k}
\right|
+\log\left|
\frac{\sqrt{\Delta E^2-m^2}\omega+\Delta E k}
     {\sqrt{\Delta E^2-m^2}\omega-\Delta E k}
\right|
\Bigg\}
\nonumber
\,,
\\&&\phantom{Spacer} \nonumber\\
J_2 \!\!\!&=&\!\!\! \int 
\frac{k^2 dk}{(\Delta E -\omega)^2\omega^4}
\\
\!\!&=&\!\!
\frac{1}{\Delta E^2}\Bigg[ 
                          \frac{1}{2m}\arctan\Big(\frac{k}{m}\Big)
                      -   \frac 12 \frac{k}{\omega^2}
                    \Bigg]
 + \frac{k}{\Delta E^3(\Delta E -\omega)}
 - \frac{2}{\Delta E^3}\frac{k}{\omega}
 - \frac{3m}{\Delta E^4}\arctan\Big(\frac{k}{m}\Big)
\nonumber\\
&+&\frac{2\Delta E^2-3m^2}{\Delta E^4\sqrt{\Delta E^2-m^2}}\frac 12
\Bigg\{
\log\left|
\frac{\sqrt{\Delta E^2-m^2}+k}{\sqrt{\Delta E^2-m^2}-k}
\right|
+\log\left|
\frac{\sqrt{\Delta E^2-m^2}\omega+\Delta E k}
     {\sqrt{\Delta E^2-m^2}\omega-\Delta E k}
\right|
\Bigg\}
\nonumber
\,,
\\&&\phantom{Spacer} \nonumber\\
J_3 \!\!\!&=&\!\!\! \int 
\frac{k^2 dk}{(\Delta E -\omega)^2\omega^6}
\\
\!\!&=&\!\!
\frac{1}{\Delta E^2}\Bigg[ 
                          \frac{1}{8m^3}\arctan\Big(\frac{k}{m}\Big)
                      +   \frac18 \frac{k}{m^2\omega^2}
                      -   \frac14 \frac{k}{\omega^4}
                    \Bigg]
 + \frac{1}{\Delta E^3}\frac23\frac{k^3}{m^2\omega^3}
 + \frac{1}{\Delta E^4}\Bigg[
                             \frac{3}{2m}\arctan\Big(\frac{k}{m}\Big)
                          -\frac{3}{2}\frac{k}{\omega^2}
                       \Bigg]
\nonumber\\
&-& \frac{1}{\Delta E^5}\frac{4k}{\omega}
 - \frac{5m}{\Delta E^6}\arctan\Big(\frac{k}{m}\Big)
 + \frac{1}{\Delta E^5}\frac{k}{(\Delta E-\omega)}
\nonumber\\
&+&\frac{4\Delta E^2-5m^2}{\Delta E^6\sqrt{\Delta E^2-m^2}}\frac 12
\Bigg\{
\log\left|
\frac{\sqrt{\Delta E^2-m^2}+k}{\sqrt{\Delta E^2-m^2}-k}
\right|
+\log\left|
\frac{\sqrt{\Delta E^2-m^2}\omega+\Delta E k}
     {\sqrt{\Delta E^2-m^2}\omega-\Delta E k}
\right|
\Bigg\}
\nonumber
\,.
\end{eqnarray}

We evaluate the above integrals at their boundaries and obtain
\begin{eqnarray}
{[I_1]}_0^\infty\!\!\!&=&\!\!\!
-\frac{1}{\Delta E}-\frac{\pi}{2}\frac{m}{\Delta E^2}
+\frac{\sqrt{\Delta E^2-m^2}}{\Delta E^2}
\frac 12 \log\left|
\frac{\sqrt{\Delta E^2-m^2}+\Delta E}{\sqrt{\Delta E^2-m^2}-\Delta E}
\right|\,,
\\&&\phantom{Spacer} \nonumber\\
{[I_2]}_0^\infty\!\!\!&=&\!\!\!
 \frac{1}{\Delta E}\frac{1}{3 m^2}
+\frac{1}{\Delta E^2}\frac{\pi}{4}\frac{1}{m}
 - \frac{1}{\Delta E^3}
-\frac{\pi}{2}\frac{m}{\Delta E^4}
+\frac{\sqrt{\Delta E^2-m^2}}{\Delta E^4}
\frac 12 \log\left|
\frac{\sqrt{\Delta E^2-m^2}+\Delta E}{\sqrt{\Delta E^2-m^2}-\Delta E}
\right|\nonumber\,,
\\&&\phantom{Spacer} \nonumber\\
{[I_3]}_0^\infty\!\!\!&=&\!\!\!
\frac{2}{15}\frac{1}{\Delta E}\frac{1}{m^4}
+\frac{\pi}{16}\frac{1}{\Delta E^2}\frac{1}{m^3}
+\frac{1}{\Delta E^3}\frac{1}{3m^2}
+\frac{\pi}{4}\frac{1}{\Delta E^4}\frac{1}{m}
-\frac{1}{\Delta E^5}
-\frac{\pi}{2}\frac{m}{\Delta E^6}
\nonumber\\
&+&\frac{\sqrt{\Delta E^2-m^2}}{\Delta E^6}
\frac 12 \log\left|
\frac{\sqrt{\Delta E^2-m^2}+\Delta E}{\sqrt{\Delta E^2-m^2}-\Delta E}
\right|\nonumber\,,
\\&&\phantom{Spacer} \nonumber\\
{[J_1]}_0^\infty\!\!\!&=&\!\!\!
-\frac{1}{\Delta E}-\frac{\pi}{2}\frac{m}{\Delta E^2}
-\frac{m^2}{\Delta E^2\sqrt{\Delta E^2-m^2}}
\frac 12 \log\left|
\frac{\sqrt{\Delta E^2-m^2}+\Delta E}{\sqrt{\Delta E^2-m^2}-\Delta E}
\right|\,,
\\&&\phantom{Spacer} \nonumber\\
{[J_2]}_0^\infty\!\!&=&\!\!
 \frac{\pi}{4}\frac{1}{\Delta E^2}\frac{1}{m}
 - \frac{3}{\Delta E^3}
 -\frac{3\pi}{2}\frac{m}{\Delta E^4}
+\frac{2\Delta E^2 -3m^2}{\Delta E^4\sqrt{\Delta E^2-m^2}}
\frac 12 \log\left|
\frac{\sqrt{\Delta E^2-m^2}+\Delta E}{\sqrt{\Delta E^2-m^2}-\Delta E}
\right|
\nonumber\,,
\\&&\phantom{Spacer} \nonumber\\
{[J_3]}_0^\infty\!\!\!&=&\!\!\!
\frac{\pi}{16}\frac{1}{\Delta E^2}\frac{1}{m^3}
+\frac 23 \frac{1}{\Delta E^3}\frac{1}{m^2}
+\frac{3\pi}{4}\frac{1}{\Delta E^4 m}
-\frac{5}{\Delta E^5}
-\frac{5\pi}{2}\frac{m}{\Delta E^6}
\nonumber\\
&+&\frac{4\Delta E^2 -5m^2}{\Delta E^6\sqrt{\Delta E^2-m^2}}
\frac 12 \log\left|
\frac{\sqrt{\Delta E^2-m^2}+\Delta E}{\sqrt{\Delta E^2-m^2}-\Delta E}
\right|
\nonumber\,.
\end{eqnarray}
When expanded up to second order in $1/\Delta E$, the above expressions
read
\begin{eqnarray}
{[I_1]}_0^\infty\!\!\!&\simeq&\!\!\!
-\frac{1}{\Delta E} -\frac{\pi}{2}\frac{m}{\Delta E^2}
+\frac{1}{\Delta E}\log\left|\frac{2\Delta E}{m}\right|
\,,
\\
{[I_2]}_0^\infty\!\!\!&\simeq&\!\!\!
\frac 13 \frac{1}{\Delta E m^2} +\frac{\pi}{4}\frac{1}{\Delta E^2 m}
\,,
\\
{[I_3]}_0^\infty\!\!\!&\simeq&\!\!\!
\frac{2}{15} \frac{1}{\Delta E m^4} +\frac{\pi}{16}\frac{1}{\Delta E^2 m^3}
\,,
\\
{[J_1]}_0^\infty\!\!\!&\simeq&\!\!\!
-\frac{1}{\Delta E} -\frac{\pi}{2}\frac{m}{\Delta E^2}
\,,
\\
{[J_2]}_0^\infty\!\!\!&\simeq&\!\!\!
\frac{\pi}{4}\frac{1}{\Delta E^2 m}
\,,
\\
{[J_3]}_0^\infty\!\!\!&\simeq&\!\!\!
\frac{\pi}{16}\frac{1}{\Delta E^2 m^3}
\,.
\end{eqnarray}

The integrals which we need for obtaining the Lamb shift in Rindler space are
($k=\sqrt{\lambda^2-m^2}$)
\begin{eqnarray}
\label{LambRindlerInt:Amp}
R_1\!\!&=&\!\!
\int \frac{dk}{\lambda^5}
\frac{1}{\Delta E - \lambda}
\\
&=& \frac{1}{\Delta E}\bigg(\frac23\frac{k}{m^4\lambda}
         +\frac13\frac{k}{m^2\lambda^3}\bigg)
+\frac{1}{\Delta E^2}\bigg(\frac{1}{2m^3}\arctan\Big(\frac{k}{m}\Big)
                           +\frac12\frac{k}{m^2\lambda^2}\bigg)
\nonumber\\
&+&\frac{1}{\Delta E^3}\frac{k}{m^2\lambda}
+\frac{1}{\Delta E^4}\frac{1}{m}\arctan\Big(\frac{k}{m}\Big)
\nonumber\\
&+&\frac{1}{\Delta E^4\sqrt{\Delta E^2-m^2}}\frac 12
\Bigg\{
\log\left|
\frac{\sqrt{\Delta E^2-m^2}+k}{\sqrt{\Delta E^2-m^2}-k}
\right|
+\log\left|
\frac{\sqrt{\Delta E^2-m^2}\lambda+\Delta E k}
     {\sqrt{\Delta E^2-m^2}\lambda-\Delta E k}
\right|
\Bigg\}
\\
{[R_1]}_0^\infty \!\!&=&\!\!
 \frac{1}{\Delta E}\frac23\frac{1}{m^4}
+\frac{1}{\Delta E^2}\frac{\pi}{4}\frac{1}{m^3}
+\frac{1}{\Delta E^3}\frac{1}{m^2}
+\frac{\pi}{2}\frac{1}{\Delta E^4}\frac{1}{m}
\nonumber\\
&+&\frac{1}{\Delta E^4\sqrt{\Delta E^2-m^2}}\frac 12
\log\left|
\frac{\sqrt{\Delta E^2-m^2}+\Delta E}
     {\sqrt{\Delta E^2-m^2}-\Delta E}
\right|
\nonumber\,,
\\&&\phantom{Spacer} \nonumber\\
\label{LambRindlerInt:Omega}
R_2\!\!&=&\!\!
\int \frac{dk}{\lambda^6}
\frac{1}{\left(\Delta E-\lambda\right)^2}
\\
&=& \frac{1}{\Delta E^2}
        \bigg(\frac38\frac{1}{m^5}\arctan\Big(\frac{k}{m}\Big)
         + \frac38\frac{k}{m^4\lambda^2}
         +\frac14\frac{k}{m^2\lambda^4}
        \bigg)
+\frac{2}{\Delta E^3}\bigg(\frac{2}{3}\frac{k}{m^4\lambda}
                        + \frac13\frac{k}{m^2\lambda^3}
                     \bigg)
\nonumber\\
&+&\frac{3}{\Delta E^4}\bigg(
                             \frac{1}{2m^3}\arctan\Big(\frac{k}{m}\Big)
                          +  \frac{k}{2m^2\lambda^2}
                       \bigg)
 +\frac{4}{\Delta E^5}\frac{k}{m^2\lambda}
 +\frac{1}{\Delta E^6}\frac{5}{m}\arctan\Big(\frac{k}{m}\Big)
\nonumber\\
&+&\frac{1}{\Delta E^5(\Delta E^2-m^2)}\frac{k}{\Delta E - \lambda}
\nonumber\\
&+&\frac{6\Delta E^2 - 5m^2}{\Delta E^6(\Delta E^2-m^2)^{3/2}}\frac 12
\Bigg\{
\log\left|
\frac{\sqrt{\Delta E^2-m^2}+k}{\sqrt{\Delta E^2-m^2}-k}
\right|
+\log\left|
\frac{\sqrt{\Delta E^2-m^2}\lambda+\Delta E k}
     {\sqrt{\Delta E^2-m^2}\lambda-\Delta E k}
\right|
\Bigg\}
\\
{[R_2]}_0^\infty 
&=&\frac{1}{\Delta E^2}
        \frac{3\pi}{16}\frac{1}{m^5}
+\frac{1}{\Delta E^3}\frac{4}{3}\frac{1}{m^4}
+\frac{1}{\Delta E^4}\frac{3\pi}{4}\frac{1}{m^3}
+\frac{1}{\Delta E^5}\frac{4}{m^2}
 +\frac{1}{\Delta E^6}\frac{5\pi}{2}\frac{1}{m}
-\frac{1}{\Delta E^5(\Delta E^2-m^2)}
\nonumber\\
&+&\frac{6\Delta E^2 - 5m^2}{\Delta E^6(\Delta E^2-m^2)^{3/2}}\frac 12
\log\left|
\frac{\sqrt{\Delta E^2-m^2}+\Delta E}
     {\sqrt{\Delta E^2-m^2}-\Delta E}
\right|
\nonumber\,.
\end{eqnarray}

\end{document}